\documentclass[reprint,showpacs,onecolumn,aps,superscriptaddress]{revtex4-2}
\usepackage[left]{lineno}
\usepackage{amsfonts}
\usepackage{amsmath}
\usepackage{amssymb}
\usepackage{graphicx}
\usepackage{verbatim}
\usepackage{textcomp}
\usepackage{upgreek}
\usepackage{threeparttable}
\usepackage{float}
\usepackage{color}
\providecommand{\U}[1]{\protect\rule{.1in}{.1in}}
\usepackage{titlesec}
\titleformat{\section}
  {\normalfont\Large\bfseries}{\thesection}{1em}{}
\titlespacing*{\section}
  {0pt}{\baselineskip}{0\baselineskip}
\titleformat{\subsection}
  {\normalfont\bfseries}{\thesection}{1em}{}
\titlespacing*{\subsection}
  {0pt}{\baselineskip}{0\baselineskip}
\usepackage{indentfirst}
\usepackage{float}

\begin{document}
\title{Picotesla-sensitivity microcavity optomechanical magnetometry}

\author{Zhi-Gang Hu$^{1,3,\dag}$, Yi-Meng Gao$^{1,3,\dag}$, Jian-Fei Liu$^{1,3,\dag}$, Hao Yang$^{1,3}$, Min Wang$^{1,3}$, Yuechen Lei$^{1,3}$, Xin Zhou$^{1,3}$, Jincheng Li$^{1,2}$, Xuening Cao$^{1,3}$, Jinjing Liang$^{1,3}$, Chao-Qun Hu$^{1,3}$, Zhilin Li$^{1,4}$, Yong-Chang Lau$^{1,3,\dagger\dagger}$, Jian-Wang Cai$^{1,3,**}$, and Bei-Bei Li$^{1,3,4,*}$\\
$^1$ Beijing National Laboratory for Condensed Matter Physics, Institute of Physics, Chinese Academy of Sciences, Beijing 100190, China\\
$^2$ School of Physics, Beihang University, Beijing 100191, China\\
$^3$ University of Chinese Academy of Sciences, Beijing 100049, China\\
$^4$ Songshan Lake Materials Laboratory, Dongguan 523808, Guangdong, China\\
$\dag$ These authors contributed equally to this work.\\
$\dagger\dagger$ yongchang.lau@iphy.ac.cn; $**$ jwcai@iphy.ac.cn; $*$ libeibei@iphy.ac.cn
}


\begin{abstract}

Cavity optomechanical systems have enabled precision sensing of magnetic fields, by leveraging the optical resonance-enhanced readout and mechanical resonance-enhanced response. Previous studies have successfully achieved scalable and reproducible microcavity optomechanical magnetometry (MCOM) by incorporating Terfenol-D thin films into high-quality ($Q$) factor whispering gallery mode (WGM) microcavities. However, the sensitivity was limited to 585~pT/Hz$^{1/2}$, over 20 times inferior to those using Terfenol-D particles. In this work, we propose and demonstrate a high-sensitivity and scalable MCOM approach by sputtering a FeGaB thin film onto a high-$Q$ SiO$_2$ WGM microdisk. Theoretical studies are conducted to explore the magnetic actuation constant and noise-limited sensitivity by varying the parameters of the FeGaB film and SiO$_2$ microdisk. Multiple magnetometers with different radii are fabricated and characterized. By utilizing a microdisk with a radius of 355~\textmu m and a thickness of 1~\textmu m, along with a FeGaB film with a radius of 330~\textmu m and a thickness of 1.3~\textmu m, we have achieved a remarkable peak sensitivity of 1.68~pT/Hz$^{1/2}$ at 9.52~MHz. This represents a significant improvement of over two orders of magnitude compared with previous studies employing sputtered Terfenol-D film. Notably, the magnetometer operates without a bias magnetic field, thanks to the remarkable soft magnetic properties of the FeGaB film. Furthermore, as a proof-of-concept, we have demonstrated the real-time measurement of a pulsed magnetic field simulating the corona current in a high-voltage transmission line using our developed magnetometer. These high-sensitivity magnetometers hold great potential for various applications, such as magnetic induction tomography and corona current monitoring.
\end{abstract}

\maketitle


\section*{Introduction}

High-sensitivity magnetometers play a crucial role in both fundamental research and practical applications, including spin physics in condensed matter systems \cite{zhao2012sensing}, dark matter detection \cite{safronova2018search}, nuclear magnetic resonance \cite{savukov2013magnetic}, magnetocardiography \cite{xiao2023movable}, magnetoencephalography \cite{pizzo2019deep}, mineral exploration \cite{edelstein2007advances}, nondestructive inspection \cite{li2023nondestructive}, magnetic induction tomography \cite{wickenbrock2014magnetic,wickenbrock2016eddy}, and current detection \cite{hagh2017wideband,lopez2019fiber,chen2014detection,xin2016development,yuan2012development}. Currently, the state-of-the-art magnetometers are based on superconducting quantum interference devices (SQUIDs), with exceptional sensitivity at the fT/Hz$^{1/2}$ level \cite{tschirhart2021imaging}. However, their reliance on cryogenic systems has resulted in elevated operational costs and limited applications. Recently, various magnetometers that can work at room temperature have been developed, such as atomic magnetometers \cite{dang2010ultrahigh}, diamond nitrogen-vacancy color center magnetometers \cite{maze2008nanoscale}, and magnetostrictive material magnetometers \cite{bucholtz1989high}. In the past few years, high-quality ($Q$) factor whispering gallery mode (WGM) optical microcavities have been widely explored for various applications due to the enhanced light-matter interactions, including quantum optics \cite{2021OE}, integrated photonic network \cite{2021S}, optical filters \cite{lei2022fully}, optical frequency combs \cite{2022NP}, and highly-sensitivity sensing \cite{2022LiuEmer,jin2021hetero,zhi2017single,tang2021omnidirectional,bei2014raman,yu2022optofluid}. Microcavity optomechanical magnetometry (MCOM) \cite{forstner2012cavity,forstner2014ultrasensitive,yu2016optomechanical,li2018quantum,li2020ultrabroadband,zhu2017polymer,colombano2020ferromagnetic,li2018invited,gotardo2023waveguide} has recently been developed by incorporating magnetostrictive materials into high-$Q$ WGM optical microcavities, which offer advantages of miniaturization and low power consumption; ease of on-chip integration; high sensitivity and broad bandwidth. In these magnetometers, the magnetic field induces strain in the magnetostrictive material, which can drive the mechanical motion of the microcavity and result in a change in the radius of the microcavity (Fig. \ref{fig1}a). This radius change shifts the optical resonance, resulting in a periodic change in the intracavity field (Fig. \ref{fig1}b). Consequently, the magnetic field can be optically read out, with its sensitivity enhanced by both the optical and mechanical resonances. Furthermore, cavity optomechanical systems have enabled precision sensing of various physical quantities, including displacement \cite{schliesser2008high}, mass \cite{yu2016cavity}, force \cite{buchmann2016complex}, acceleration \cite{krause2012high,guzman2014high}, and ultrasound waves \cite{yang2022high,yang2023micropascal,Tang2023acoustic,meng2022dissi}, etc.

The first MCOM was realized by manually depositing a particle of magnetostrictive material, Terfenol-D, on the top of a microcavity and bonding it using epoxy \cite{forstner2012cavity}. However, its sensitivity was limited to a few hundreds of nT/Hz$^{1/2}$ due to the small overlap between the magnetostrictive force and the mechanical mode of the microcavity. The sensitivity was then improved by depositing the Terfenol-D particle into the central hole of the microcavity \cite{forstner2014ultrasensitive,li2018quantum,li2020ultrabroadband}, achieving a peak sensitivity of 26 pT/Hz$^{1/2}$ at 10.5~MHz \cite {li2020ultrabroadband}. Resonant magnon-assisted optomechanical magnetometers have recently been realized, achieving a peak sensitivity of 850~pT/Hz$^{1/2}$ at a frequency of 206~MHz \cite{colombano2020ferromagnetic}. These magnetometers working at MHz frequency band have various potential applications, such as corona current monitoring in high-voltage transmission line systems \cite{xin2016development}, magnetic induction tomography \cite{wickenbrock2014magnetic,wickenbrock2016eddy}, and nuclear magnetic resonance in the human body \cite{savukov2013magnetic}. However, the fabrications of these magnetometers involving manual deposition of magnetostrictive materials are intricate, posing significant challenges for scalable and reproducible fabrication.  A more scalable and reproducible fabrication method was developed by sputtering coating Terfenol-D thin films into SiO$_2$ microtoroids \cite{li2018invited}. Nevertheless, the achieved peak sensitivity of 585~pT/Hz$^{1/2}$ is more than one order of magnitude inferior to that achieved in magnetometers using Terfenol-D particles \cite{li2020ultrabroadband}. Additionally, the vulnerability to oxidation of Terfenol-D further limits their applications.

In this work, we have successfully developed scalable and high-sensitivity MCOM by sputter coating thin films of a new magnetostrictive material, FeGaB, onto high-$Q$ microdisk cavities. We have made improvements in two crucial aspects, leading to improved performance compared to previous work. Firstly, we replace the magnetostrictive material Terfenol-D with FeGaB, as FeGaB exhibits superior soft magnetic properties, including lower coercivity, lower saturation magnetic field, and a larger maximum piezomagnetic coefficient \cite{dong2018characterization, lou2007soft, lou2009giant}. Secondly, microdisk cavities are used in this work, offering improved scalability and reproducibility compared to microtoroids used in earlier studies \cite{li2018invited}. We conduct comprehensive theoretical studies to explore the relationship between sensitivity and the geometric parameters of the FeGaB films and SiO$_2$ microdisks. Subsequently, we fabricate magnetometers with varying microdisk radii and measure their sensitivities within the frequency range of 5~kHz to 30~MHz. Remarkably, we achieve a peak sensitivity of 1.68~pT/Hz$^{1/2}$ at 9.52~MHz using a magnetometer with a SiO$_2$ microdisk radius of 355~\textmu m and a thickness of 1~\textmu m, along with a FeGaB film radius of 330~\textmu m and a thickness of 1.3~\textmu m. This sensitivity surpasses previous scalable MCOM by over two orders of magnitude. Furthermore, we explore a proof-of-concept application by detecting a pulsed magnetic field signal, simulating corona current detection in high-voltage transmission lines. These findings demonstrate the potential of our fabricated high-sensitivity magnetometers for various practical applications, such as magnetic induction tomography and corona current detection.

\section*{Results}

\subsection*{\textbf{Theoretical analysis}}

\begin{figure}[h]
\centering
\includegraphics[width=12cm]{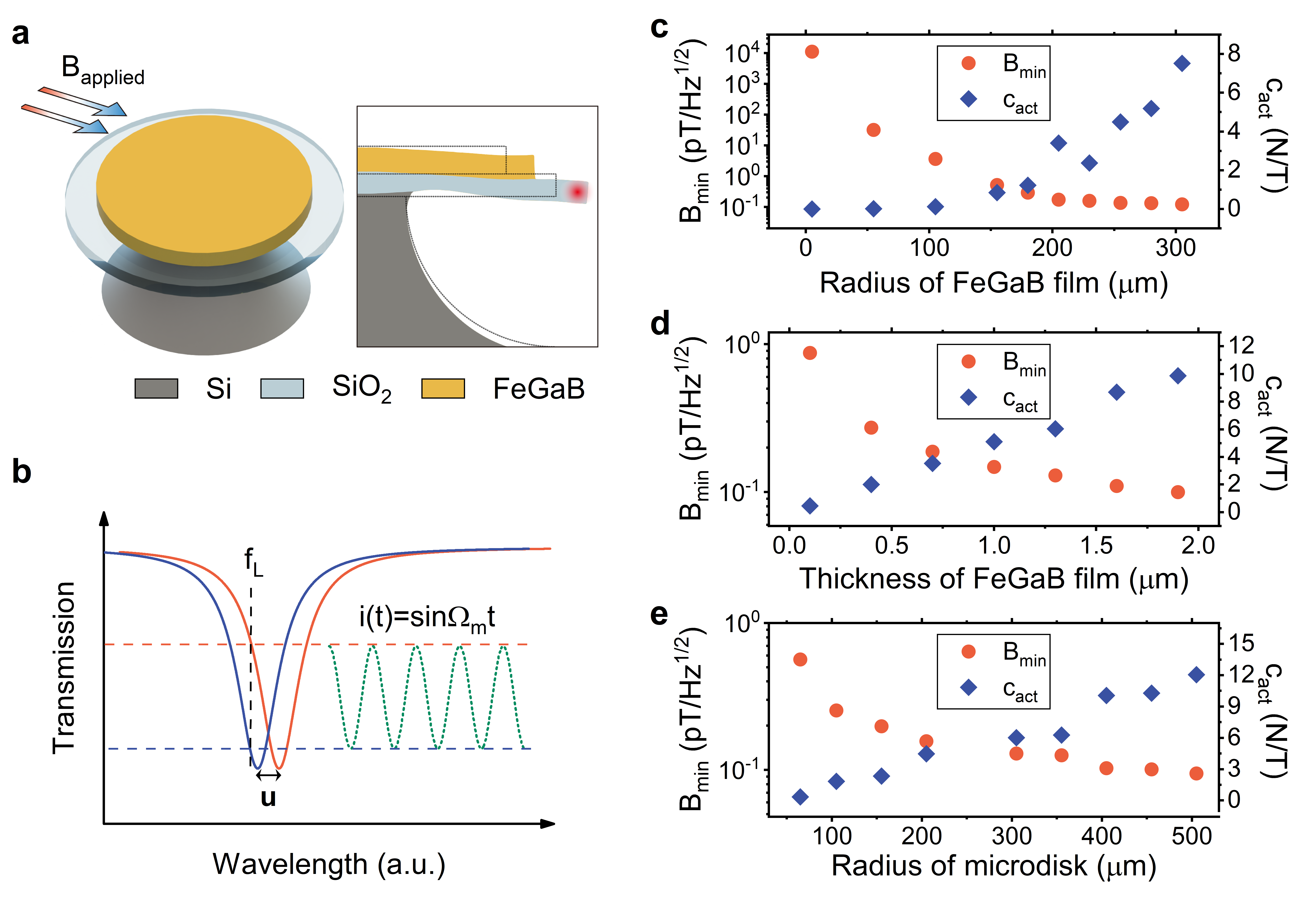}
\caption{Designed structure and theoretical analysis of the MCOM. \textbf{a} Schematic structure (left panel) and the cross-section (right panel) of the magnetometer. The optical WGM is confined along the inner surface. The RBM motion changes the microcavity circumference, thereby shifting the optical resonance. \textbf{b} Optical readout principle. The magnetostriction-induced mechanical displacement \textbf{\emph{u}} shifts the optical transmission spectrum from the blue curve to the orange one periodically. When the laser frequency $f_{L}$ is locked on the side of the optical resonance, the frequency shift can be converted into an intensity modulation of the detected photocurrent $i(t)$ with the mechanical angular frequency of $\Omega_\mathrm{m}$. \textbf{c}-\textbf{e} Simulation results for magnetic actuation constants $c_{\mathrm{act}}$ (blue diamonds) and sensitivity $B_{\mathrm{min}}$ (orange dots) for the RBM of the magnetometer, as a function of the radius of FeGaB film \textbf{c}, the thickness of FeGaB film \textbf{d}, and the radius of microdisk \textbf{e}, respectively.}
\label{fig1}
\end{figure}

Figure \ref{fig1}a presents a schematic structure of our designed MCOM. It consists of a SiO$_{2}$ microdisk with a thin FeGaB film deposited on top, supported by a silicon pedestal. The magnetometer supports both high-$Q$ optical and mechanical modes, as illustrated in the right panel of Fig. \ref{fig1}a. When a magnetic field is applied, the magnetostriction of the FeGaB film can drive the mechanical displacement (\textbf{\emph{u}}) of the microcavity, which modifies the circumference of the microcavity and therefore shifts the optical resonance, as illustrated in Fig. \ref{fig1}b. The magnetic field amplitude can thus be converted into the amplitude or phase modulation of the intracavity field, which can be optically read out from the detected photocurrent $i(t)$.


We first conduct a thorough theoretical analysis of the sensitivity of our MCOM. To read out the magnetic field, we employ a technique where the laser frequency is locked on the side of the optical resonance using thermal locking \cite{mcrae2009thermo}. This allows for a straightforward implementation of the intensity modulation readout mechanism (as shown in Fig. \ref{fig1}b), which is commonly used in optomechanical sensing experiments \cite{yang2022high,yang2023micropascal,li2020ultrabroadband,gotardo2023waveguide}. In the case of the intensity modulation readout mechanism, it is necessary to consider the detuning between the laser frequency $\omega_L$ and the cavity resonance frequency $\omega_\mathrm{o}$: $\Delta=\omega_\mathrm{L}-\omega_\mathrm{o}$. However, previous studies have determined the power spectral densities of displacement noise under the assumption of zero frequency detuning \cite{bowen2015quantum,aspelmeyer2014cavity,Bowen2017}. Hence, we take this detuning into account and obtain the sensitivity of the magnetometers using the intensity modulation readout mechanism under the critical-coupled condition (see section 1 in the Supplementary Information). The sensitivity of the magnetometer, $B_{\mathrm{min}}(\Omega)$, defined as the minimum detectable magnetic field at different frequencies with a measurement resolution bandwidth (RBW) of 1~Hz, can be expressed as:

\begin{equation}
\begin{split}
   B_{\mathrm{min}}(\Omega)
   &=\frac{1}{c_{\mathrm{act}}}\sqrt{(\frac{S_{\mathrm{thermal}}}{A_1|\chi_\Delta|^2}+\frac{S_{\mathrm{shot}}}{A_1|\chi_\Delta|^2}+\frac{S_\mathrm{back-action}}{A_1|\chi_\Delta|^2})\cdot 2p_{\mathrm{zp}}^2} \\
   &=\frac{1}{c_{\mathrm{act}}}\sqrt{2m_{\mathrm{eff}}\gamma k_B \mathcal{T}+\frac{p_{\mathrm{zp}}^2}{A_1|\chi_\Delta|^2}+2A_2m_{\mathrm{eff}}\gamma\hbar\Omega_\mathrm{m}|C_{\mathrm{eff}}^\Delta|},
\end{split}
\label{eq1}
\end{equation}
where $\Omega$ is the Fourier frequency, $A_1=\frac{4g^2\kappa_1[\Delta^2\cos^2{\theta}+(\kappa_1^2+\Omega^2)\sin^2{\theta}-\kappa_1\Delta \sin{2\theta}]}{(\kappa_1^2-\Omega^2+\Delta^2)^2+4\kappa_1^2\Omega^2}$, $A_2=\frac{(\kappa_1^2+\Omega^2)(\Delta^2+\kappa_1^2+\Omega^2)}{(\kappa_1^2-\Omega^2+\Delta^2)^2+4\kappa_1^2\Omega^2}$, are constants associated with the frequency detunig $\Delta$. Here, $\kappa_1$ represents the external energy decay rate of the optical mode, and $\theta=\mathrm{arctan}(\frac{\mathrm{Im}(\alpha_\mathrm{out})}{\mathrm{Re}(\alpha_\mathrm{out})})$ denotes the phase angle of the output field. $\mathrm{Re}(\alpha_\mathrm{out})$ and $\mathrm{Im}(\alpha_\mathrm{out})$ denote the real and imaginary parts of the cavity field, respectively. $g = g_\mathrm{0}\sqrt{N}$ is the cavity-enhanced optomechanical coupling rate, where $g_\mathrm{0}$ is the single-photon optomechanical coupling rate defined as $g_0=G_{\mathrm{om}}x_\mathrm{zp}$. Here, $G_{\mathrm{om}}=\frac{d\omega_\mathrm{o}}{dR}$ is the optomechanical coupling coefficient, which quantifies how much $\omega_\mathrm{o}$ shifts with the radius ($R$) change. $x_\mathrm{zp}$ denotes the zero-point displacement of the mechanical mode. $N$ represents the steady-state photon number in the cavity. $m_\mathrm{eff}$ is the effective mass of the mechanical mode. $\gamma$ is the energy decay rate of the mechanical mode. $k_B$ is the Boltzmann constant, $\mathcal{T}$ is the temperature. $\Omega_\mathrm{m}$ is the angular frequency of the mechanical mode. $C_{\mathrm{eff}}^\Delta(\Omega)=\frac{4g^\mathrm{2}}{\kappa\gamma(1-2i\Omega/\kappa)^\mathrm{2}}$ is effective optomechanical cooperativity. $\chi_\Delta(\Omega)=\frac{\Omega_\mathrm{m}}{-\Omega^\mathrm{2}-i\Omega\gamma+\Omega_\mathrm{m}^\mathrm{2}+\frac{4g^\mathrm{2}\Omega_\mathrm{m}\Delta}{(\kappa/2-i\Omega)^\mathrm{2}+\Delta^\mathrm{2}}}$ is the modified mechanical susceptibility in the presence of the optical field. $p_{\mathrm{zp}}=\sqrt{\frac{\hbar m_\mathrm{eff}\Omega_\mathrm{m}}{2}}$ denotes the zero-point-fluctuation momentum. $c_{\mathrm{act}}=F/B$ is the magnetic actuation constant, which characterizes how well the magnetic field is converted into an applied force on the mechanical resonator \cite{forstner2012cavity}.

According to Eq. (\ref{eq1}), the sensitivity is determined by the sensor noise and the value of $c_{\mathrm{act}}$. The sensor noise is represented by the formula inside the square root, with the three terms representing thermal noise, shot noise, and back-action noise, respectively. The back-action noise is typically much smaller than the thermal noise and therefore is negligible (see section 1 in the Supplementary Information). In comparison to previous studies those are under the condition of zero frequency detuning \cite{aspelmeyer2014cavity,Bowen2017}, the thermal-noise-dominant sensitivity remains unchanged, which is determined by $\mathcal{T}$, $m_{\mathrm{eff}}$, and $\gamma$. Conversely, the shot noise-dominant sensitivity is modified and is dependent on the detuning $\Delta$. When the laser is blue-detuned ($\Delta > 0$), $\gamma$ is reduced, which is beneficial for magnetic field sensing. Conversely, when the laser is red-detuned ($\Delta < 0$), $\gamma$ is increased, which can deteriorate the sensitivity. Additionally, the value of $\Delta$ also affects $A_1$. To quantify the contribution of the shot noise to the total noise, we investigate the ratio of thermal noise to shot noise ($S_{\mathrm{thermal}}/S_{\mathrm{shot}}$) as a function of the laser detuning. It is found that $S_{\mathrm{thermal}}/S_{\mathrm{shot}}$ reaches its maximum when $\Delta=\frac{\kappa}{2\sqrt{3}}$, which corresponds to a transmission $T=1/4$, as shown in Fig. S1b. (For details, see section 1 in Supplementary Information.) Consequently, in our following experiment of magnetic field sensing, we lock the laser frequency on the blue side of the optical mode and keep the transmission at 1/4. 

To enhance the magnetic field sensitivity, it is crucial to optimize $c_{\mathrm{act}}$ to be as high as possible. Specifically, the expression of $c_{\mathrm{act}}$ in the frequency domain can be expressed as \cite{yu2018modelling}:

\begin{equation}
\begin{split}
c_{\mathrm{act}}(\Omega)=\frac{F_\mathrm{signal}(\Omega)}{B_\mathrm{applied}(\Omega)}=\frac{\mathrm{max}[\textbf{\emph{u}}(\Omega)]\Omega_m m_{\mathrm{eff}}}{\chi(\Omega)B_{\mathrm{applied}}(\Omega)},
\end{split}
\label{eq2}
\end{equation}
where $\chi=\frac{\Omega_\mathrm{m}}{-\Omega^\mathrm{2}-i\Omega\gamma+\Omega_\mathrm{m}^\mathrm{2}}$ is the mechanical susceptibility, and $\mathrm{max}[\textbf{\emph{u}}(\Omega)]$ is the maximum displacement of the mechanical mode induced by an applied magnetic field, obtained using the finite element method (FEM) simulation. The value of $c_{\mathrm{act}}(\mathrm{\Omega})$ depends on the geometric parameters of the magnetometer and the piezomagnetic coefficient of the magnetostrictive film. Therefore, we simulate $c_{\mathrm{act}}(\mathrm{\Omega})$ and $B_{\mathrm{min}}(\Omega_\mathrm{m})$ with different geometric parameters using the method proposed in Ref. \cite{yu2018modelling}. Specifically, we optimize the parameters for the RBM, since its motion is primarily radial, which aligns well with the direction of magnetostriction. Furthermore, the RBM exhibits the highest optomechanical coupling coefficient $G_{\mathrm{om}}$, resulting in the highest optical transduction sensitivity.

Figure \ref{fig1}c illustrates the relationship between $c_{\mathrm{act}}$ ($B_{\mathrm{min}} (\Omega_\mathrm{m})$) and the radius of the FeGaB film. In these simulations, we keep the radius of the microdisk at 305~\textmu m and the radius of the silicon pedestal at 106~\textmu m, which are consistent with our experimental parameters. When the radius of the FeGaB film is smaller than that of the silicon pedestal, the deformation induced by the magnetostriction effect of the film is constrained by the pedestal, leading to a relatively small value of $c_{\mathrm{act}}$. As the radius of the FeGaB film exceeds that of the silicon pedestal, the mechanical modes can be effectively actuated by the magnetostrictive force, resulting in a larger $c_{\mathrm{act}}$. Therefore, a larger radius of FeGaB film indicates a better magnetic field sensitivity. Taking into account the absorption of optical field energy by FeGaB material and the alignment precision in the fabrication process, we set the radius of FeGaB film to be 25~\textmu m smaller than that of the microdisk. In Fig. \ref{fig1}d, we study the effect of the thickness of FeGaB film on $c_{\mathrm{act}}$ and $B_{\mathrm{min}} (\Omega_\mathrm{m})$, where the radii of the microdisk and FeGaB film are set as 305~\textmu m and 280~\textmu m, respectively. The results suggest that thicker FeGaB films yield better sensitivity and magnetic actuation constants. However, as the thickness increases, there is a higher risk of the FeGaB film peeling off from the microdisk. As a compromise, we utilize a thickness of 1.3~\textmu m for the FeGaB film in the fabrication of our magnetometers. In Fig. \ref{fig1}e, we vary the radius of the microdisk from 65~\textmu m to 505~\textmu m, while keeping the radius of FeGaB film 25~\textmu m smaller than that of the microdisk. It shows that the theoretical sensitivity for each radius maintains below 1~pT/Hz$^{1/2}$, with a slight improvement as the radius of the microdisk increases. This result shows that by adjusting the microdisk radius, we can achieve highly sensitive detection of magnetic fields across a wide frequency range in the megahertz bands.

\subsection*{\textbf{Device fabrication and characterization}}

\begin{figure}[H]
\centering
\includegraphics[width=10cm]{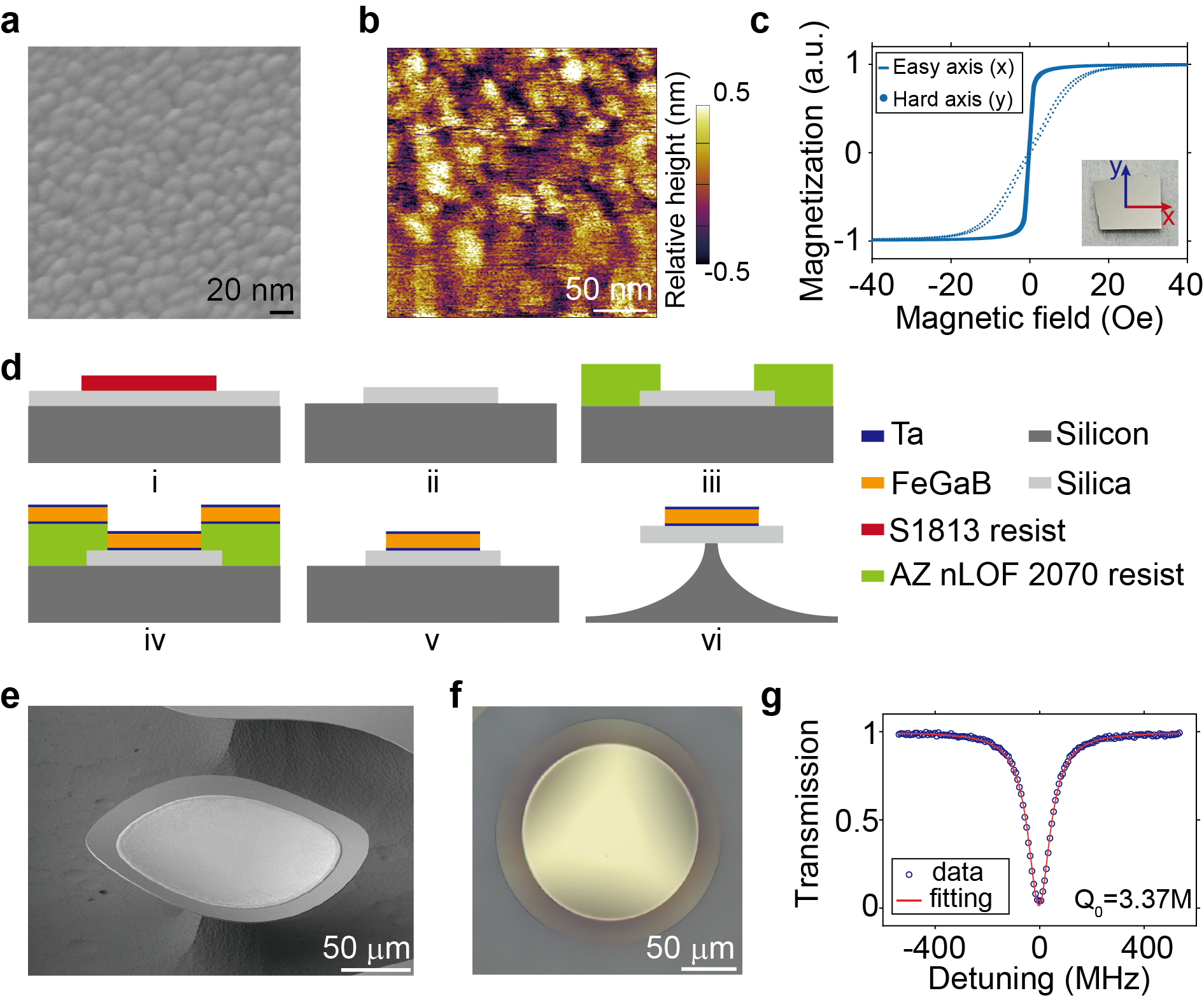}
\caption{Fabrication and characterization of the magnetometers. \textbf{a} Scanning electron microscope (SEM) image of the FeGaB film. \textbf{b} Atomic force microscope (AFM) of a 500~nm $\times$ 500~nm region of the FeGaB film. RMS roughness of the FeGaB film is $\sim$ 540~pm. \textbf{c} In-plain magnetic hysteresis loops measured using a vibrating sample magnetometer (VSM). The lines and scatters correspond to the loops measured along the easy-axis (x) and hard-axis (y) directions respectively. \textbf{d} Schematic of the fabrication process of the magnetometers. \textbf{e}-\textbf{f} SEM and optical microscope images of the fabricated magnetometer, respectively. The slight warping of the SiO$_2$ microdisk outer edge is due to the stress within the FeGaB film. \textbf{g} Optical transmission spectrum of the fabricated magnetometer, showing an intrinsic $Q$ factor of around 3.37$\times10^6$ for the WGM of the microdisk, in the 1550~nm band under the critical-coupled condition.}
\label{fig2}
\end{figure}

To characterize the properties of the magnetostrictive material, we start by sputter coating the FeGaB film onto a 1~cm$\times$1~cm blank silica substrate with the deposition details shown in the Materials and methods section. Figure \ref{fig2}a presents a scanning electron microscope (SEM) image of the deposited Ta/FeGaB/Ta stack film. This suggests that the incorporation of boron atoms into the FeGa alloys results in a refined grain size \cite{lou2007soft}. Figure \ref{fig2}b presents the two-dimensional surface tomography obtained with atomic force microscope (AFM). The root-mean-square (RMS) roughness of this film is determined to be approximately 540~pm. The magnetization of the FeGaB film is measured using a vibrating sample magnetometer (VSM, Lake Shore 7404). To determine the saturation magnetization and assess the magnetic anisotropy of the films, in-plane hysteresis loops are recorded with field parallel and perpendicular to the field applied during the deposition process, denoted as x and y in the inset of Fig. \ref{fig2}c. Figure \ref{fig2}c displays the hysteresis loops of the deposited FeGaB film in both the easy-axis (x) and hard-axis (y) directions, represented in the solid curve and dots, respectively. The FeGaB film exhibits an anisotropy field of approximately 20~Oe and a coercivity of around 0.4~Oe. The saturation magnetization is calculated to be around 1.08-1.23$\times$10$^3$~emu/cm$^3$, which is consistent with the values reported in Ref. \cite{lou2007soft}. Notably, our deposited FeGaB film has significantly improved soft magnetic properties compared with Terfenol-D, which has an anisotropy field of 1000~Oe and a coercivity of approximately 6~Oe \cite{hathaway1995magnetomechanical,mech2017influence}. Additionally, although the saturation magnetostriction coefficient of Terfenol-D ($\sim$2000~ppm) is significantly larger than that of FeGaB ($\sim$70~ppm), the maximum piezomagnetic coefficient of FeGaB ($\sim$6~ppm/Oe) is higher than that of Terfenol-D ($\sim$2.6~ppm/Oe) \cite{lou2009giant,dong2018characterization}. These parameters indicate that FeGaB is more sensitive to small magnetic fields compared with Terfenol-D.

Figure \ref{fig2}d illustrates the fabrication process of our designed magnetometers, with the fabrication details shown in the Materials and methods section. Figures \ref{fig2}e and f display the SEM and optical microscope images of the fabricated magnetometers with a microdisk radius of 105~\textmu m, respectively. The slight warping observed in the SiO$_2$ microdisk is attributed to the stress within the FeGaB film, which is measured to be approximately 260~MPa. Figure \ref{fig2}g presents the optical transmission spectrum of the microdisk, showing an intrinsic optical $Q_0$ factor of $3.37\times10^6$ under the critical-coupled condition.

\subsection*{\textbf{Sensitivity}}
\begin{figure}[h!]
\centering
\includegraphics[width=10cm]{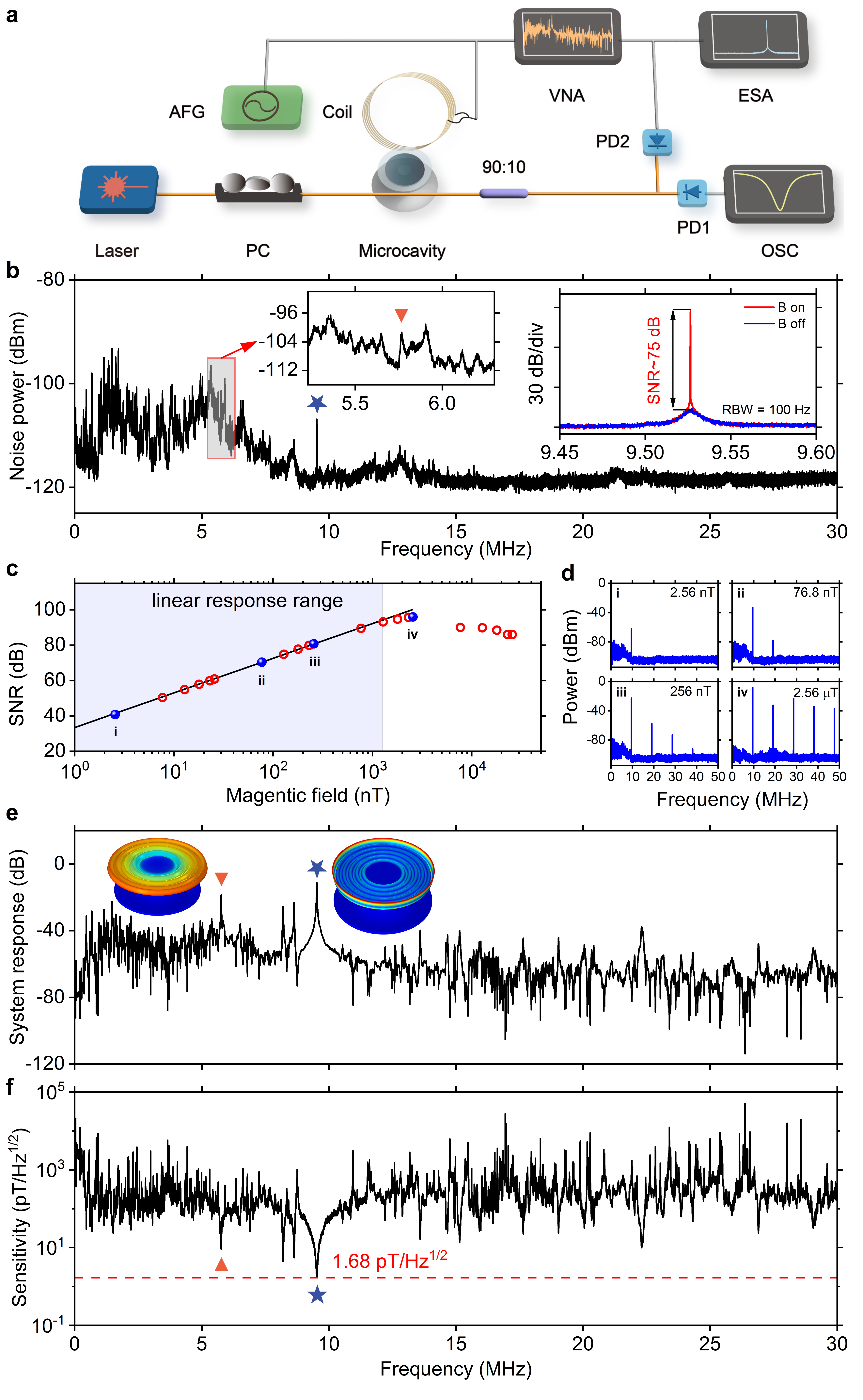}
\caption{Sensitivity measurement of the magnetometers. \textbf{a} Schematic of the experimental setup. AFG: arbitrary function generator, PC: polarization controller, PD: photodetector, VNA: vector network analyzer, ESA: electronic spectrum analyzer, OSC: oscilloscope. \textbf{b} Noise power spectrum of a magnetometer with a radius of 355~\textmu m, with the peaks representing the thermally excited mechanical modes. Left inset: zoomed-in noise power spectrum within the frequency range of 5.2~MHz-6.3~MHz. Right inset: power spectra with (red) and without (blue) the magnetic field in the frequency range of 9.45~MHz-9.6~MHz. The applied AC magnetic field is around 128~nT at 9.53~MHz. \textbf{c} Signal-to-noise ratio (SNR) at 9.53~MHz as a function of increasing AC magnetic field. A linear response of SNR is fitted in the black line using data within the shaded region. A nonlinearity emerges when the magnetic field exceeds 1.28~\textmu T. \textbf{d} Power spectra obtained at the magnetic fields of 2.56~nT, 76.8~nT, 256~nT, and 2.56~\textmu T, respectively, orderly identified by blue dots in \textbf{c}. \textbf{e} System response of the magnetometer in the frequency range of 5~kHz-30~MHz. Insets show the simulated displacement profiles of RBM and HOFM using FEM simulation. \textbf{f} Sensitivity spectrum in the frequency range of 5~kHz-30~MHz, highlighting a peak sensitivity of approximately 1.68~pT/Hz$^{1/2}$ at 9.52~MHz. In \textbf{b}, \textbf{e}, and \textbf{f}, the orange triangles and blue stars denote the RBM and HOFM, respectively.}
\label{fig3}
\end{figure}

The sensitivity of our magnetometers is characterized using the setup depicted in Fig. \ref{fig3}a. A tunable laser in the 1550~nm band is utilized to evanescently couple light into the WGM of the microdisk via a tapered fiber. The polarization of the light is optimized using a fiber polarization controller to match the polarization of the WGM. The transmitted light from the tapered fiber is collected by a photodetector (PD) and subsequently measured using an oscilloscope (OSC), electronic spectrum analyzer (ESA), and vector network analyzer (VNA), respectively. The transmission spectrum of the microdisk is recorded using the OSC as the laser wavelength is scanned. A coil positioned 3~mm away from the magnetometer is driven by an AFG or VNA, to generate a magnetic field along the hard-axis direction of the FeGaB film. The magnetic field within the frequency range of 0-50~MHz is calibrated using a commercial Gaussmeter and the VNA. The wavelength of the laser is thermally locked on the blue-detuned side of the high-Q optical mode. The mechanical motion induced by the applied magnetic field modulates the radius of the microdisk, thereby altering its optical resonance. This is converted into a periodic modulation in the intensity of the transmitted light. To optimize the readout sensitivity, the laser frequency detuning is set as $\Delta=\frac{\mathrm{\kappa}}{2\sqrt{3}}$ under the critical-coupled condition, corresponding to a transmission of $T=1/4$. The noise power spectrum of the mechanical mode is measured by the ESA. Additionally, the magnetic response of the magnetometer at different frequencies can be obtained by sweeping the frequency using the VNA.

Figure \ref{fig3}b shows the measured noise power spectrum $S(\Omega)$ in the frequency range of 5~kHz-30~MHz of a magnetometer with a microdisk radius of 355~\textmu m. The multiple characteristic peaks correspond to thermally excited mechanical modes. The left inset in Fig. \ref{fig3}b presents the zoomed-in noise power spectrum around the RBM. By applying a magnetic field of around 128~nT at the frequency of $\Omega_{\mathrm{ref}}/2\pi = 9.53~$MHz, a signal peak above the thermal noise is obtained, exhibiting a signal-to-noise ratio (SNR) of 75~dB, as shown in the right inset of Fig. \ref{fig3}b. To validate that the applied magnetic field of 128~nT falls within the linear response range of the magnetometer, we investigate the SNR at 9.53~MHz, under increasing the magnetic field ranging from 2.56~nT to 25.6~\textmu T. Figure \ref{fig3}c presents the SNR at 9.53~MHz for different magnetic fields, demonstrating a linear response range of 2.56~nT-1.28~\textmu T, with its linear fitting shown in the black line using data within the shaded region. Consequently, the sensitivity at 9.53~MHz can be obtained from the following equation \cite{forstner2012cavity}:

\begin{equation}
    B_{\min}(\Omega_{\mathrm{ref}})=\frac{|B_{\mathrm{applied}}(\Omega_{\mathrm{ref}})|}{\sqrt{\mathrm{SNR}(\Omega_{\mathrm{ref}})\Delta\Omega_{\mathrm{RBW}}}},
    \label{eq}
\end{equation}
where $B_{\mathrm{applied}}(\Omega_{\mathrm{ref}})$ represents the applied magnetic field at 9.53~MHz and $\Delta\Omega_{\mathrm{RBW}} = 2\mathrm{\pi}\times100$~Hz is the RBW of the ESA. The sensitivity at 9.53~MHz can be calculated as 2.28~pT/Hz$^{1/2}$. As the AC magnetic field exceeds 1.28~\textmu T,  a noticeable nonlinearity in the response emerges. Figure \ref{fig3}d presents the power spectra in the range of 5~kHz to 50~MHz at four different magnetic fields marked by blue dots in Fig. \ref{fig3}c. It is shown that with the increase of the magnetic field, higher-order mechanical sidebands arise due to the nonlinear transduction. Since the cavity mode is a Lorentzian lineshape, the optical readout signal for displacement is a harmonic oscillation only for a small displacement \cite{yang2022high}. After the saturation of the first-order sideband, the dynamic range can be effectively expanded by detecting SNRs of the higher-order mechanical sidebands \cite{javid2021cavity}. Additionally, a decline of SNR is observed as the applied magnetic field increases to 7.68~\textmu T. This decline can be attributed to the reduction in intracavity power when the mechanical displacement induced by the magnetic field is very large \cite{hu2021generation,schliesser2008resolved,krause2015nonlinear}. In the case of large displacement, cascaded $n$ photon-phonon scattering events occur. This leads to a large fraction of the intracavity photons being scattered, resulting in additional transmission dips near each detuning $\Delta=n\Omega_\mathrm{m}$. The transmission spectra at different magnetic fields are presented in section 2 in the Supplementary Information.

\begin{figure}[h!]
\centering
\includegraphics{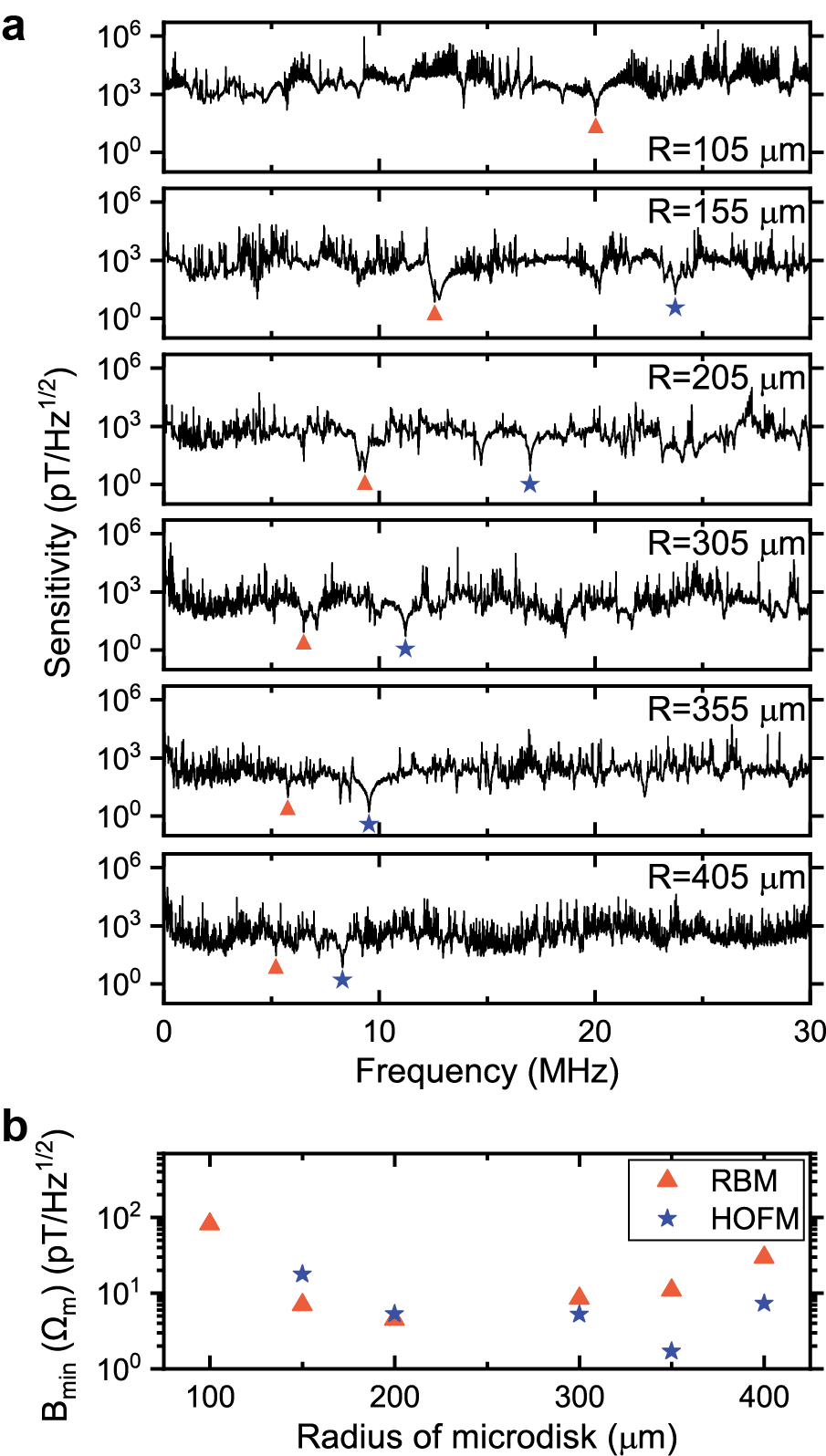}
\caption{\textbf{a} Sensitivity spectra of different-sized magnetometers within the frequency range of 5~kHz-30~MHz. The microdisk radii are 105~\textmu m, 155~\textmu m, 205~\textmu m, 305~\textmu m, 355~\textmu m and 405~\textmu m from top to bottom. The dips marked by orange triangles correspond to the RBMs, located at 20.02~MHz, 12.57~MHz, 9.34~MHz, 6.49~MHz, 5.77~MHz, and 5.20~MHz from top to bottom. The dips marked by blue stars correspond to the HOFMs, located at 23.73~MHz, 17.00~MHz, 11.21~MHz, 9.52~MHz, and 8.30~MHz from top to bottom. \textbf{b} Optimal sensitivity $B_{\mathrm{min}}$ of the RBMs (orange triangles) and HOFMs (blue stars), as a function of the microdisk radius.}
\label{fig4}
\end{figure}

The measured system response $N(\Omega)$ in the frequency range of 5~kHz to 30~MHz is presented in Fig. \ref{fig3}e. The orange triangle and blue star represent the RBM at 5.76~MHz and the high-order flapping mode (HOFM) at 9.53~MHz, whose displacement profiles are shown in the insets, obtained by FEM simulation. Using the equation \cite{forstner2012cavity,yang2022high}

\begin{equation}
B_{\min}(\Omega)=B_{\min}(\Omega_{\mathrm{ref}})\frac{B_{\mathrm{applied}}(\Omega)}{B_{\mathrm{applied}}(\Omega_{\mathrm{ref}})}\sqrt{\frac{N(\Omega)}{N(\Omega_{\mathrm{ref}})}\cdot\frac{S(\Omega_{\mathrm{ref}})}{S(\Omega)}},
\label{eq4}
\end{equation}
we calculate the sensitivity in a broad frequency range, as shown in Fig. \ref{fig3}f. The orange triangle and blue star in Fig. \ref{fig3}f correspond to the RBM and HOFM modes indicated in the insets of Fig. \ref{fig3}e. In Eq. (\ref{eq4}), $S(\Omega)$ is the noise power spectrum obtained without any applied magnetic field, as shown in Fig. \ref{fig3}b. $B_{\mathrm{applied}}(\Omega_{\mathrm{ref}})$ denotes the amplitude of the applied magnetic field within the frequency range. Note that the amplitude of the magnetic field varies with frequency when the amplitude of the driving voltage is the same, due to the impedance change of the coil at different frequencies caused by inductance. The peak sensitivity of 1.68~pT/Hz$^{1/2}$ is achieved at a frequency of 9.52~MHz, consistent with the theoretical sensitivity of 1.5~pT/Hz$^{1/2}$. However, the sensitivity of the RBM at 5.76~MHz is 9.18~pT/Hz$^{1/2}$, which is significantly lower than the theoretical prediction of 126~fT/Hz$^{1/2}$. The discrepancy in $B_{\mathrm{min}}$ for the RBM may be ascribed to the coupling between the RBM and other mechanical modes. As illustrated in the left inset of Fig. \ref{fig3}b, the power spectrum of the RBM overlaps significantly with that of other mechanical modes, indicating that magnetic fields may simultaneously drive multiple mechanical modes around the RBM, thereby weakening the drive to the RBM. Notably, the achieved sensitivity is more than two orders of magnitude improvement compared with previous scalable and reproducible magnetometers \cite{li2018invited}. Even when compared with the magnetometers that used Terfenol-D particles \cite{li2020ultrabroadband}, our device still demonstrates a factor of 15 times sensitivity enhancement.  

To investigate the relationship between the sensitivity of the magnetometers and the microdisk radius, we fabricate magnetometers with various radii. Figure \ref{fig4}a presents typical sensitivity spectra of six magnetometers with radii of 105~\textmu m, 155~\textmu m, 205~\textmu m, 305~\textmu m, 355~\textmu m, and 405~\textmu m, respectively, within the frequency range of 5~kHz-30~MHz. The magnetic response of the magnetometer can be significantly enhanced by the mechanical resonances of microdisks, resulting in numerous dips at mechanical resonance frequencies throughout the magnetic sensitivity spectra. The dips marked by orange triangles and blue stars represent the RBMs and HOFMs, respectively. The HOFM of the magnetometer with the microdisk radius of 105~\textmu m is not marked because its mode frequency exceeds 30~MHz. It is observed that the frequencies of both mechanical modes decrease as the radius of the microdisk increases. Figure \ref{fig4}b illustrates the optimal sensitivity $B_{\mathrm{min}}$ of the RBM and HOFM for microdisks with different radii. For the RBM, the optimal sensitivity improves as the radius of the microdisk increases from 105~\textmu m to 205~\textmu m, which is consistent with the theoretical results in Fig. \ref{fig1}e. However, this trend deviates from the theory for microdisks with radii of 305~\textmu m to 405~\textmu m. This may be attributed to the degradation of the mechanical quality factors of microdisks with larger radii. It is caused by the increasing clamping loss associated with thicker silicon pedestals in microdisks with larger radii to prevent the structure from cracking due to the stress in the FeGaB film. A similar trend of optimal sensitivity is also observed for the HOFM.

\subsection*{\textbf{Corona current detection}}

\begin{figure}[htbp]
\centering
\includegraphics{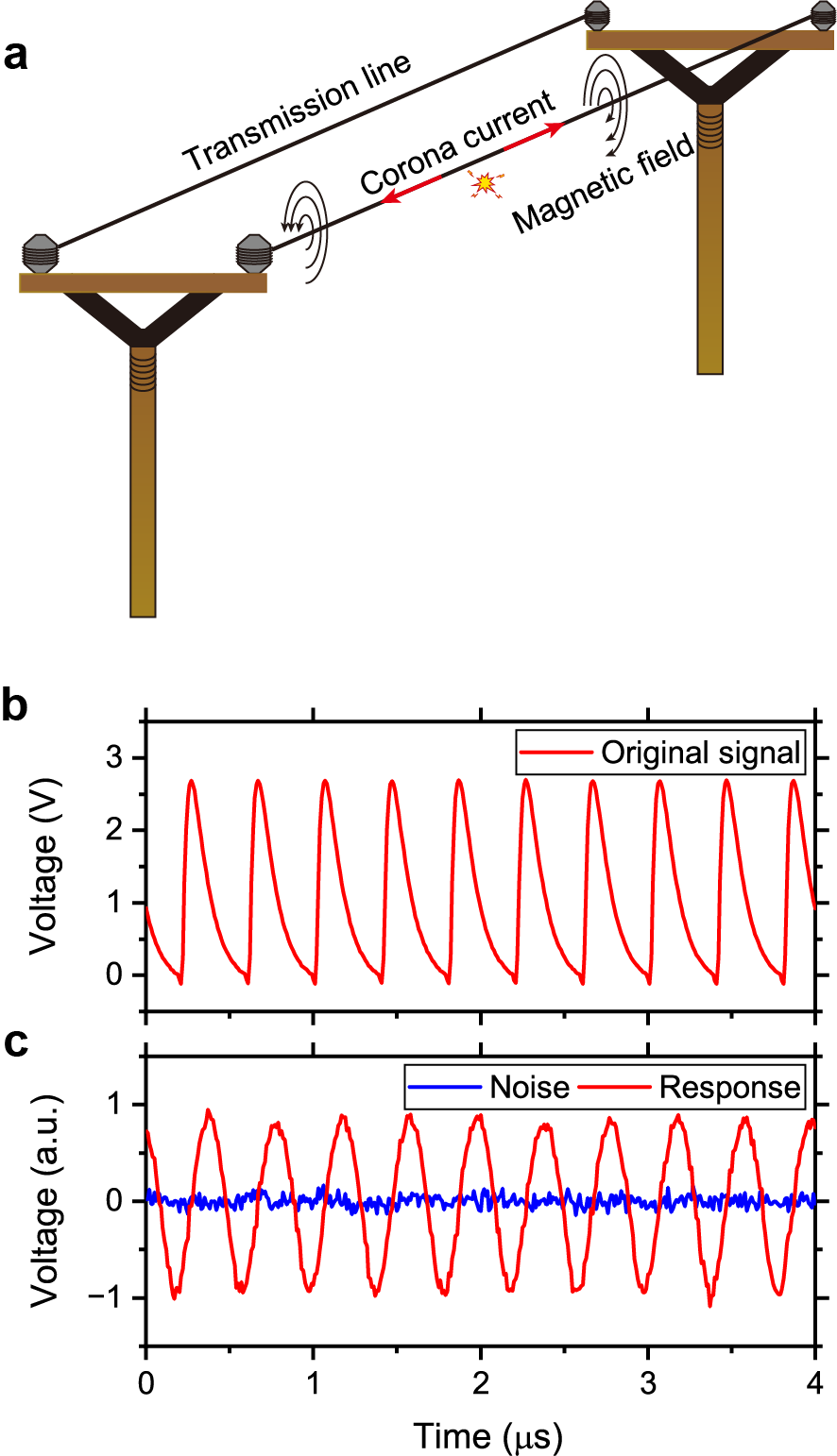}
\caption{ \textbf{a} Schematic of corona current monitoring through magnetic field detection. \textbf{b} Original voltage signal generated by the AFG to simulate the corona current. \textbf{c} Response of the magnetometer with a microdisk radius of 205~\textmu m. The red (blue) curve shows the AC component of the optical transmission with (without) the signal.}
\label{fig5}
\end{figure}
High voltage direct current (HVDC) transmission lines play an essential role in long-distance power transmission. When the high voltage is applied, the intense electric field exits near the conductor's surface. Once the electric field exceeds the threshold of air ionization, self-sustaining corona discharge occurs in the surrounding air, which results in energy loss along with audible noise, release of ozone gas, and radio interference. To mitigate these effects, one can improve the design of the HVDC transmission lines, by analyzing the characteristics of the corona current induced by the discharge. The corona current has rich frequency components typically within the range of 0-100~MHz \cite{xin2016development}. Consequently, detecting corona current requires sensors with high sensitivity and broad bandwidth. In recent years, several photoelectric-based methods for corona current measurement have been proposed. Most of these methods involve placing resistors in series with the transmission line to record the electric current signals, which are then read out using photodiodes \cite{wang2008measurement,xin2016development}. In this work, we demonstrate corona current measurement using our magnetometers, by detecting the weak magnetic field generated by the corona current outside the transmission line, as illustrated in Fig. \ref{fig5}a. Compared with the previously reported photoelectric-based method, the MCOM provides a non-destructive and all-optical measurement method. It has the advantage of reduced insulation requirement and immunity to electromagnetic interference.





In our proof-of-concept application, we simulate the corona current on a 500~kV transmission line using an AFG to generate a pulsed voltage signal, as shown in Fig. \ref{fig5}b. This pulsed voltage signal is applied to the coil to generate a pulsed magnetic field, that can be measured using our magnetometer. The repetition frequency is 2.5~MHz, and the peak amplitude is 2.822~V, corresponding to a peak amplitude of 31.695~mA in the coil \cite{xin2016development,wang2018pulse}. This current generates a pulsed magnetic field with a peak amplitude of 11.4~\textmu T at the position of the magnetometer. The red curve in Fig. \ref{fig5}c shows the time-domain response of the magnetometer to the pulsed current signal, with a microdisk radius of 205~\textmu m, while the blue curve represents the noise background without the pulsed current signal. In these results, the DC component of the optical transmission of the magnetometer has been filtered out. The measured response of the magnetometer exhibits excellent consistency with the original signal, with a slight distortion in the waveform, which originates from the varied cavity response for magnetic fields of different frequencies (see Fig. \ref{fig3}e). By performing a fast Fourier transform to both the measured response and noise, the time-domain response can be converted to the frequency domain, as displayed in Fig. S3 in the Supplemental Information. It can be seen that the corona current has a wide frequency distribution. With an RBW of 1~kHz, our magnetometer can measure the corona current at 2.5~MHz with an amplitude SNR of 30~dB (see section 3 in Supplementary Information). This demonstrates the potential of our magnetometers for corona current characterization in the real application scenario.

\section*{Discussion}

In conclusion, we propose and fabricate a new type of scalable MCOM that exhibits ultrahigh sensitivity, by sputter-coating a thin film of magnetostrictive material FeGaB onto the SiO$_2$ microdisk cavity. Through theoretical analysis, we derive the expression for photocurrent detected by the photodetector and noise-limited sensitivity using the intensity modulation readout mechanism, considering the laser frequency detuning. To optimize the sensitivity, we study the relationship between the sensitivity of the magnetometers and the geometric parameters of the FeGaB film and the SiO$_2$ microdisk. Our theoretical analysis reveals that larger radii of FeGaB film and SiO$_2$ microdisk, as well as a thicker FeGaB film, result in a higher magnetic actuation constant and improved sensitivity. We experimentally fabricate multiple magnetometers with different radii (105~\textmu m, 155~\textmu m, 205~\textmu m, 305~\textmu m, 355~\textmu m, and 405~\textmu m) and characterize their sensitivities. We achieve an impressive minimum detectable magnetic field of 1.68~pT/Hz$^{1/2}$ at 9.52~MHz, which is two orders of magnitude better than previous scalable magnetometers \cite{li2018invited}. Finally, we simulate the corona current in the high-voltage transmission line and successfully detect the pulsed magnetic field generated by this current using the fabricated magnetometers. 

Utilizing the FeGaB material in our magnetometers to convert magnetic field signals into measurable mechanical signals represents a significant breakthrough. The peak sensitivity achieved in this work is the highest reported thus far in MCOM, making FeGaB a promising material for enhancing magnetometer sensitivity. Additionally, exploiting the $\Delta E$ effect (Young's modulus change in the presence of strain) \cite{spetzler2021exchange} of the FeGaB film can further improve sensitivity in the low-frequency range. Further developments can aim towards fully integrated MCOM using silicon nitride ring resonators incorporated with FeGaB films. These high-sensitivity magnetometers open up new possibilities for various applications, such as corona current detection and magnetic induction tomography.

\section*{Materials and methods}
\subsection*{\textbf{Deposition of the FeGaB film}}
To characterize the magnetic properties of magnetostrictive material, we sputter the FeGaB film onto a blank silica substrate. As a first step, a 5~nm-thick Ta layer is deposited to improve the adhesion between the FeGaB film and the silica substrate. Subsequently, a layer of FeGaB with a thickness of 1~\textmu m is sputtered using a target with a composition of (Fe$_{80}$Ga$_{20}$)$_{88}$B$_{12}$. A radio frequency (RF) power of 100~W yields a sputtering rate of 0.13~nm/s with an Ar pressure of 0.5~Pa. During the deposition, a static magnetic field of about 300~Oe is applied in the plane of the FeGaB film to induce a uniaxial in-plane anisotropy. Finally, a 5~nm-thick Ta cap layer is sputtered on top of the FeGaB film to prevent it from oxidation. 

\subsection*{\textbf{Fabrication of the magnetometer}}
The magnetometer is fabricated as follows. Starting with a silicon wafer with a 1-\textmu m-thick thermally oxidized silica on top, we pattern the SiO$_2$ microdisks using photolithography with a positive photoresist S1813 (step i). The pattern is then transferred to the oxide layer through inductively coupled plasma (ICP) etching utilizing a CHF$_3$/Ar/H$_2$ chemistry (step ii). Subsequently, a negative photoresist AZ nLOF 2070 is spin-coated onto the wafer, and a second photolithography process is performed to define the region for FeGaB deposition while protecting the perimeter of each disk (step iii). A stacked layer of Ta/FeGaB/Ta films, with thicknesses of 5~nm/1.3~\textmu m/5~nm, is sputter coated on the surface of the SiO$_2$ microdisk (step iv). The excess Ta and FeGaB on the photoresist are removed through a lift-off process using Remover PG (step v). Finally, the silicon substrate is etched using XeF$_2$ to form a silicon pedestal, leaving the SiO$_2$ microdisk suspended (step vi). In the XeF$_2$ etching process, the FeGaB film is protected by the Ta cap layer from reacting with XeF$_2$. 

\section*{Acknowledgment}

The authors thank Professor Yun-Feng Xiao and Professor Warwick Bowen for useful discussions and acknowledge the funding from the National Key Research and Development Program of China (2021YFA1400700, 2021YFB3501400); National Natural Science Foundation of China (NSFC) (62222515, 12174438, 11934019, 12274438); Basic frontier science research program of Chinese Academy of Sciences (ZDBS-LY-JSC003); CAS Project for Young Scientists in Basic Research (YSBR-100). Zhilin Li is grateful for the support from the Youth Innovation Promotion Association of the Chinese Academy of Sciences(No. 2021008). This work is also supported by the Micro/nano Fabrication Laboratory of Synergetic Extreme Condition User Facility (SECUF).

\section*{Author contributions}

B.-B.L. conceived the idea. Z.-G.H., Y.-M.G., and J.-F.L. fabricated the devices and implemented the measurements. J.-W.C., Y.-C.L., C.-Q.H., J.L., and Z.L. helped with the sputter coating of the FeGaB films. Z.-G.H. and J.-F.L. performed the numerical simulations. B.-B.L., Z.-G.H., Y.-M.G., and J.-F.L. discussed the results and wrote the manuscript. All authors participated in the manuscript revision, reviewed the final draft, and gave their approvals for submission.

\section*{Data availability}

The data that support the plots within this paper and other findings of this study are available from the corresponding authors upon request.

\section*{Conflict of interest}

The authors declare no competing interests.

\section*{Supplementary information}  

The online version contains supplementary material available at

\bibliography{ref}

\end{document}